\documentclass[aps,twocolumn,groupedaddress,showpacs,floatfix]{revtex4}

\usepackage{amssymb,amsmath}
\usepackage{graphicx}
\usepackage[english]{babel}
\usepackage{color}

\begin{document}

\title{Rubidium 87 Bose-Einstein condensate in an optically plugged quadrupole trap}

\author{R. Dubessy, K. Merloti, L. Longchambon, P.-E. Pottie, T. Liennard, A. Perrin, V. Lorent and H. Perrin\footnote{helene.perrin@univ-paris13.fr}}

\affiliation{Laboratoire de physique des lasers, CNRS and Universit\'e Paris 13, 99 avenue J.-B. Cl\'ement, F-93430 Villetaneuse}

\begin{abstract}
We describe an experiment to produce $^{87}$Rb Bose-Einstein condensates in an optically plugged magnetic quadrupole trap, using a blue-detuned laser.
Due to the large detuning of the plug laser with respect to the atomic transition, the evaporation has to be carefully optimized in order to efficiently overcome the Majorana losses. We provide a complete theoretical and experimental study of the trapping potential at low temperatures and show that this simple model describes well our data. In particular we demonstrate methods to reliably measure the trap oscillation frequencies and the bottom frequency, based on periodic excitation of the trapping potential and on radio-frequency spectroscopy, respectively. We show that this hybrid trap can be operated in a well controlled regime that allows a reliable production of degenerate gases.
\end{abstract}
\pacs{67.85.-d, 37.10.Gh}
\maketitle

\section{Introduction}
\label{sec:intro}

Atom traps combining magnetic and optical potentials have been used since the achievement of the first Bose-Einstein condensates (BECs) in 1995~\cite{Davis1995}. They take advantage of the large trapping volume offered by magnetic quadrupole traps which facilitate the loading of a large atom number from magneto-optical traps (MOTs). Moreover, they allow for efficient evaporative cooling dynamics thanks to the initial linear shape of the confinement, while still offering large trapping frequencies at the end of the evaporation. Furthermore, they prevent the trapped atoms from undergoing Majorana spin flips~\cite{Majorana1932}, relying on the dipole force induced by a laser beam to push the atoms outside the low-magnetic-field region where these losses are the highest.

While solutions relying on red-detuned lasers have been demonstrated~\cite{Lin2009}, an alternative consists in using a blue-detuned laser as an optical plug at the center of the magnetic quadrupole trap; see Fig.~\ref{fig:plugged_trap}. The latter strategy has the advantage of minimizing resonant scattering processes responsible for atom losses and heating. A widespread choice for the design of the optical plug consists in using standard lasers at 532\,nm. These lasers indeed commonly reach several watts of power and have long demonstrated their stability and reliability. This design is particularly well suited for $^{23}$Na (sodium) atoms, given the vicinity of the $D$ lines at 589\,nm. It has hence recently allowed for the fast preparation of large $^{23}$Na BECs~\cite{Naik2005,Heo2011}.

Here we show that the same setup can also be used to efficiently obtain $^{87}$Rb (rubidium) BECs despite the large detuning of the optical plug laser from the $^{87}$Rb main transition at 780\,nm. Indeed, we are able to keep at all times a high collision rate while minimizing Majorana losses by modifying the trap shape during the evaporative cooling ramp. We typically obtain $2\times 10^5$ atoms in a quasi-pure BEC after a total evaporation time of 20\,s. We finally present an experimental investigation of the trap parameters and describe how to characterize the cold gas during the cooling process.

The paper is organized as follows: In Sec.~\ref{sec:BEC} we perform a thorough analysis of the optically plugged trap. In Sec.~\ref{sec:evaporation} we show how to optimize the evaporative cooling dynamics. In Sec.~\ref{sec:characterization}, we give an overview of the experimental methods allowing us to infer the trap parameters and also explain how to characterize the cold gas along the way to a BEC. Additional details on the experimental setup are given in the Appendix.

\section{Optically plugged trap}
\label{sec:BEC}

\begin{figure}[b]
\begin{center}
\includegraphics[width=\linewidth]{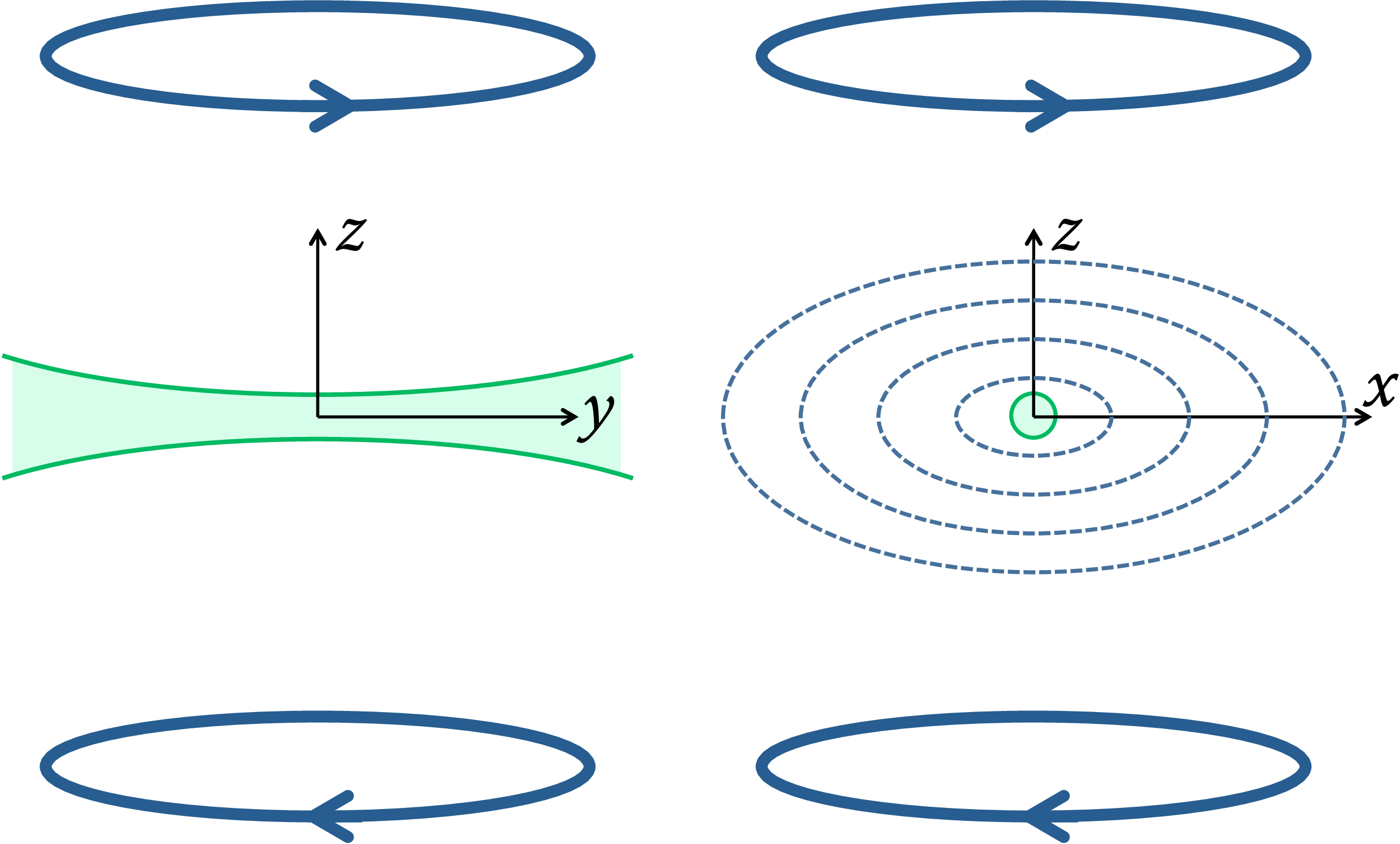}
\caption{(Color online) Principle of the optically plugged quadrupole trap. A blue-detuned focused laser beam plugs the central part of a quadrupole magnetic trap, where the field is zero. Left: The magnetic field is produced by a pair of coils of axis $z$ aligned with gravity, with opposite currents. The beam propagates along the $y$ axis. Right: The isomagnetic lines are represented by dashed lines in the $xz$ plane, together with a line of constant intensity of the plug beam, orthogonal to the plane.}
\label{fig:plugged_trap}
\end{center}
\end{figure}

In this section, we present a detailed description of optically plugged
quadrupole traps, with the geometry depicted in Fig.~\ref{fig:plugged_trap}. When the plug beam is centered exactly at the magnetic field zero, we are able to derive analytical expressions for the trap position and its corresponding oscillation frequencies, as well as to quantify the sensitivity of these quantities to the relevant experimental parameters. These results are generalized to the case of a slightly off-centered plug beam. This study allows us to infer the dependence of the trap characteristics on the experimental parameters and in turn to estimate the heating rate due to their fluctuations, in particular regarding beam pointing stability.

\subsection{Trap geometry}

As stated in the Introduction, optically plugged quadrupole traps result from the combination of a quadrupole magnetic field with a blue detuned laser beam focused near the symmetry center of the quadrupole, where the magnetic field is zero. Taking into account the gravitational field as well, the corresponding potential can be written
\begin{equation}
U(\textbf{r}) = U_B(\textbf{r})+U_D(\textbf{r})+M g z
\label{eq:potential}
\end{equation}
where $U_B(\textbf{r})$ is the magnetic component of the potential, $U_D(\textbf{r})$ its optical component, $M$ the atomic mass, and $g$ the gravity acceleration.

\begin{figure}[t]%
	\centering
	\includegraphics{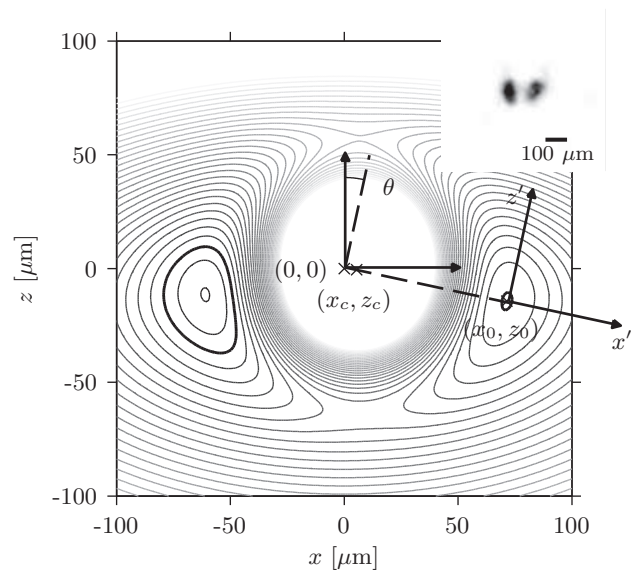}
	\caption{Calculated trapping potential in the $xz$ plane in the $F=1,m_F=-1$ ground state resulting from the combination of the quadrupole field, the optical plug centered at position $(x_c,z_c)$ and gravity, see Eq.(\ref{eq:potential}). The parameters are given in Table~\ref{tab:param}. The lines are isopotentials separated by 20\,kHz. The bold lines refer to the isopotential 1\,kHz above the right minimum at $(x_0,z_0)$. The eigenaxes of the trap $(x',z')$, making an angle $\theta$ with the reference axes, are also shown. Inset: \textit{In situ} absorption image of ultracold atoms confined in the two minima of the optically plugged trap.}%
  \label{fig:potential}
\end{figure}

Depending on the relative choice of the quadrupole magnetic field's orientation to the plug beam direction, different trap configurations are possible:  there can be a single minimum~\cite{Heo2011}, a pair of minima~\cite{Davis1995,Naik2005} or an infinite number of degenerate minima spread over a circle~\cite{Naik2005}. In the following we will consider only the case of a plug beam propagating along one of the horizontal axes of a magnetic quadrupole field of vertical symmetry axis, which corresponds to our experimental configuration (see Fig.~\ref{fig:plugged_trap}). In general such a potential presents two minima separated by two saddle points; see Fig.~\ref{fig:potential}.

As depicted in Fig.~\ref{fig:plugged_trap}, the quadrupole magnetic field is produced in our experimental setup by a pair of coils with vertical axis $z$ and can be written
\begin{equation}\label{eq:ub}
U_B(\textbf{r}) = \hbar\alpha\sqrt{x^2+y^2+4z^2}
\end{equation}
where $\alpha = \mu b'/\hbar$ and $\mu = m_F g_F \mu_B$, with $b'$ the horizontal magnetic gradient, $g_F$ the Land\'e factor in the ground state $F$, $\mu_B$ the Bohr magneton, and $m_F$ the atomic spin projection on the magnetic field axis. The potential $U_B$ is hence cylindrically symmetric around $z$.

The plug beam is produced by a green laser at 532\,nm propagating along the $y$ direction and focused to a waist $w_0$ near the zero-field position. Denoting by $(x_c,z_c)$ the focus position of the beam and neglecting the effect of the finite Rayleigh length --- typically much larger than the atomic sample --- the dipolar potential reads~\cite{Grimm2000}
\begin{equation}\label{eq:ud}
U_D(\textbf{r}) = U_0\exp\left[-2\frac{(x-x_c)^2+(z-z_c)^2}{w_0^2}\right],
\end{equation}
where $U_0$ is the maximum light shift at the beam waist. $U_0$ is proportional to the plug power $P$ and to the inverse squared waist, $U_0 \propto P/w_0^2$~\cite{note4}.

\subsection{Trap characteristics}

From Eqs.~\eqref{eq:ub} and~\eqref{eq:ud}, we see that the trap potential $U$ is completely determined by five parameters, namely, $b'$, $P$, $w_0$, and $(x_c,z_c)$. In this section, we restrict our study to the case of a centered beam, $x_c=z_c=0$. Nevertheless, our main conclusions remain valid if the plug beam is slightly off centered, the trap positions and the corresponding oscillation frequencies being only weakly affected.

Symmetries of the potential $U$ allow us to make a few general statements about the trap geometry. First of all, the dependence of $U$ on $y$ is restricted to the term $U_B$, even in $y$, which imposes the requirement that all the potential minima (and saddle points) should belong to the $y=0$ plane. Hence, the $y$ axis will always be an eigenaxis of the trap in a harmonic approximation around one of the trap minima. Second, $U$ is even in $x$.  As a consequence, the two saddle points belong to the $x=0$ plane, and the two trap minima are symmetric with respect to this plane. In the following we will hence concentrate on the minimum located on the right side $(x_0,0,z_0)$. Because of gravity, the trap depth is limited by the lower saddle point belonging to the $z$ axis, at position $(0,0,z_s)$, with $z_s<0$.

Minimizing Eq.~\eqref{eq:potential}, we can express the coordinates $x_0$ and $z_0$ as
\begin{equation}
x_0=r_0 \sqrt{1-4\varepsilon}  \quad \text{and} \quad z_0 = - r_0\sqrt{\varepsilon}.
\label{eq:x0z0}
\end{equation}
Here, we have introduced the dimensionless parameter $\varepsilon$ describing the relative strength of gravity and the magnetic gradient as $\varepsilon =\left[Mg/(3\hbar\alpha)\right]^2$. The effective radius $r_0=\sqrt{x_0^2+4z_0^2}$ can be written
\begin{equation}
r_0 = \frac{\xi}{\sqrt{1-3\varepsilon}} w_0,
\label{eq:r0}
\end{equation}
where the dimensionless parameter $\xi$ is the solution of the equation $\xi \exp[-2\xi^2]=\hbar\alpha w_0\sqrt{1-3\varepsilon}/[4U_0]$, verifying $\xi>0.5$~\cite{note1}. The isomagnetic surface $B_0$ on which the  minima of potential $U$ are located is directly linked to $r_0$ through the relation 
\begin{equation}\label{eq:B_0r_0}
\mu B_0=\hbar\alpha r_0.
\end{equation}

From Eq.~\eqref{eq:x0z0} and~\cite{note1}, we see that a few conditions must be fulfilled to ensure the existence of a minimum at position  $(x_0,0,z_0)$, namely, $\varepsilon<1/4$ and $U_0/[\hbar\alpha w_0]>\sqrt{e} \sqrt{1-3\varepsilon}/2$, where $e=\exp[1]$. The first relation is usually verified since with typical experimental parameters gravity is small compared to the magnetic field gradient. The second one requires that the light shift overcome the effect of the magnetic gradient on the size of the waist.

We now perform a harmonic approximation of the potential at the position of the minimum $(x_0,0,z_0)$ and derive the corresponding oscillation frequencies:
\begin{subequations}
\begin{eqnarray}
\omega_y &=& \sqrt{\frac{\hbar\alpha}{Mr_0}}, \label{eq:omegay}\\
\omega_{x'} &=& \sqrt{4\xi^2 - 1}\,\omega_y,  \label{eq:omegax}\\
\omega_{z'} &=& \sqrt{3(1-4\varepsilon)}\, \omega_y.  \label{eq:omegaz}
\end{eqnarray}
\label{eq:freq_osc}
\end{subequations}
The frequency $\omega_y$ appears as the natural frequency scale of the trap and  is generally the smallest one. Interestingly, from Eqs.~\eqref{eq:B_0r_0} and~\eqref{eq:omegay}, we see that $\omega_y$ is completely determined by the value of the magnetic field at the trap minimum, $B_0$, and the magnetic field gradient $b'$ through the relation $\omega_y=\sqrt{\mu b'^2/MB_0}$.

The two largest frequencies $\omega_{x'}$ and $\omega_{z'}$ correspond to the
eigenaxes $x'$ and $z'$ which are rotated with respect to the $x$ and $z$ axes
by the angle $\theta$, defined by $\tan\theta =
-\sqrt{\varepsilon/(1-4\varepsilon)} = z_0/x_0$ (see also
Fig.~\ref{fig:potential}). The angle $\theta$ is usually small. We point out that
the eigenaxis $x'$ coincides with the line linking the  center of the magnetic field $(0,0)$ to the  center of the trap $(x_0,z_0)$. In a basis centered in $(0,0)$ and rotated by $\theta$, the coordinates of the  center of the trap are $x'_0 = \sqrt{x_0^2+z_0^2}=\sqrt{1-3\varepsilon}\,r_0$ and $z'_0=0$.

The positions of the saddle points as well as the trap depth can also be
calculated analytically. The exact expression is somewhat complicated. However,
we notice that the distance of the saddle point from the  center is of the same
order as the distance $x_0$ or $r_0$. The contribution of $U_D$ is then about
the same at both positions, whereas the contribution of $U_B$ is twice as large
at the saddle point, due to the larger gradient along $z$. Including the effect of gravity, we can thus estimate the trap depth to be $\hbar\alpha r_0 -M g r_0 = \hbar\alpha r_0(1 - 3\sqrt{\varepsilon})$.

\subsection{Sensitivity to the trap parameters}
\label{sec:noise}

\subsubsection{Magnetic gradient, beam power and waist}
From Eq.~\eqref{eq:r0} and setting $\zeta = 1/(4\xi^2-1) = \omega_y^2/\omega_{x'}^2$, we can express the dependency of the effective radius $r_0$ on the experimental parameters as
\begin{equation}
r_0 \propto b'^{-\frac{\zeta+3\varepsilon}{1-3\varepsilon}}, \quad \quad r_0 \propto P^{\zeta}, \quad \quad r_0 \propto w_0^{1- 3 \zeta}.
\label{eq:r0dep}
\end{equation}
With our experimental parameters, presented in Table~\ref{tab:param}, the three exponents are $-0.26$, 0.13 and 0.6 respectively. This implies that the effective radius $r_0$ is relatively stable to fluctuations of the plug beam power and the magnetic field gradient and is only slightly sensitive to the fluctuations of the beam waist, the latter being certainly one of the most stable experimental parameters.

\begin{table}
\centering
\begin{tabular}{lccccccc}
\hline
\hline
Parameter & $b'$ & $P$ & $w_0$ & $x_c$ & $z_c$ & $U_0/k_B$&depth\\
\hline
Unit &G$\cdot$cm$^{-1}$&W&$\mu$m&$\mu$m&$\mu$m& $\mu$K& $\mu$K\\
\hline
Value & 55.4 & 5.8 & $46$ & $5.5$ & $-0.5$&100&5.3\\
\hline
Uncertainty & $\pm 0.6$& $\pm 0.1$ & $\pm 3$ & $\pm 3$ & $\pm 3$&$\pm 15$ &\\
\hline
\hline
& $\varepsilon$ & $\xi$ & $\zeta$ & $r_0$ & $\nu_{x'}$ & $\nu_y$ & $\nu_{z'}$\\
\hline
Unit &&&&$\mu$m&Hz&Hz&Hz\\
\hline
Value &  0.0337~ & ~$1.47$~ & ~0.131~ & ~77~ & ~$220$~ & ~$76.6$~ & ~$121.1$\\
\hline
Uncertainty &$\pm 0.0007$ & $\pm 0.04$ & $\pm 0.007$ & $\pm 1$ & $\pm 10$&$\pm 0.4$&$\pm 0.5$\\
\hline
\hline
\end{tabular}
\caption{Set of parameters used in the experiment, and characterized in
Sec.~\ref{sec:characterization}, leading to the isopotential lines plotted
in Fig.~\ref{fig:potential}. Atoms are prepared in the $F=1,m_F=-1$ ground state. $b'$, $P$, and the oscillation frequencies $\nu_i$ are
measured (see Sec.~\ref{sec:measOsc}); the other parameters and the related uncertainty are deduced from these measurements. Here, $\nu_i$ stands for $\omega_i/(2\pi)$, where $i=x',y,z'$.}
\label{tab:param}
\end{table}

Fluctuations in $r_0$ are responsible for fluctuations of the trap position $(x_0,z_0)$, which leads to linear heating through dipolar excitation, with a constant temperature time derivative $\dot{T}$~\cite{Gehm1998}. $\dot{T}$ is proportional to the power spectrum density (PSD) of the position noise, which can be linked to the PSD of the noise in $b'$, $P$ and $w_0$ at the trap frequencies through Eq.~\eqref{eq:r0dep}. Similarly, fluctuations in $r_0$ also lead to fluctuations in the oscillation frequencies, which produce an exponential heating due to parametric excitations, with a rate $\Gamma_{\rm param}$ proportional to the PSD at twice the trap frequencies~\cite{Gehm1998}. From the relations \eqref{eq:r0dep}, we can estimate the maximum allowed density power spectrum of relative fluctuations of $b'$, $P$ and $w_0$ that ensure a linear heating rate $\dot{T}$ and an exponential heating rate $\Gamma_{\rm param}$ below an arbitrary given threshold.

As the trap minimum always lies in the plane $y=0$, fluctuations in the parameters do not change this coordinate and the contribution of the $y$ direction to dipolar heating is zero~\cite{note2}. On the other hand, as $z'_0=0$ for a centered plug beam, the relative fluctuations of $P$ and $w_0$ do not induce dipolar heating along $z'$. Fluctuations in $b'$ do change $z'_0$ because of a change in the angle $\theta$, through $\delta z'_0 = \sqrt{x_0^2+z_0^2}\,\delta\theta$. However, even for $b'$, the dominant contribution to dipolar heating is along the direction $x'$.

For a threshold at $\dot{T}=1$\,nK$\cdot$s$^{-1}$, we find that the maximum power spectral density of relative fluctuations should be $S_{b'}(\nu_{x'})<-100$\,dB$\cdot$Hz$^{-1}$, $S_{P}(\nu_{x'})<-95$\,dB$\cdot$Hz$^{-1}$, and $S_{w_0}(\nu_{x'})<-110$\,dB$\cdot$Hz$^{-1}$ for $b'$, $P$, and $w_0$, respectively. These requirements are easily fulfilled with commercial power supplies and lasers; see the Appendix. For parametric heating, all three directions have to be taken into account. If the noise spectrum at frequencies $2\nu_i$ ($i=x',y,z'$) is the same as the one given above at $\nu_i$, the parametric heating rate due to the fluctuations in the three parameters remains very small, $\Gamma_{\rm param} < 10^{-4}$\,s$^{-1}$. Parametric heating should thus not be the dominant heating mechanism in the trap.

\subsubsection{Beam pointing}
Finally, we discuss the sensitivity of the trap to the beam pointing stability. When the plug beam is focused exactly in the  center of the magnetic trap, the two minima are symmetrically placed with respect to the $yz$ plane and present the same trap depth. However, small displacements in the plug position of the order of a few micrometers unbalance the depths and consequently the populations of the trap minima. They also modify the positions of the minima and the oscillation frequencies inside them. For large displacements, one of the minima can even disappear.

We first remark that the symmetry with respect to the $xz$ plane is still preserved with an off-centered beam, which implies that the trap minima always satisfy $y_0=0$~\cite{note2}. Moreover, the expression of the oscillation frequency $\omega_y$ as a function of the gradient $b'$ and the magnetic field $B_0$ at the trap minimum still holds. For displacements small as compared to $w_0$, we can estimate the shift induced on the position of the minimum in the $xz$ plane and on the oscillation frequencies. We find for the position of the minimum
\begin{subequations}
  \begin{eqnarray}
   \Delta x_0 &=& \frac{1-16\varepsilon/3}{1-4\varepsilon}\,x_c -\frac{4}{3}\frac{\sqrt{\varepsilon}}{\sqrt{1-4\varepsilon}}\, z_c,\\
    \Delta z_0 &=& -\frac{4}{3}\frac{\sqrt{\varepsilon}}{\sqrt{1-4\varepsilon}}\,x_c -\frac{1}{3}\,z_c.
  \end{eqnarray}
  \label{eq:plugshift}
\end{subequations}
We infer from these equations that $r_0$ is changed only by $x_c$, with a coefficient almost unity:
\begin{equation}
\Delta r_0 = \frac{1}{\sqrt{1-4\varepsilon}}x_c.
\label{eq:r0ofxc}
\end{equation}
For $x_c=0$ and $z_c\neq 0$, the trap minima stay on the isomagnetic lines defined by the value $r_0$. The eigenaxes, given by the angle $\theta$, can thus be significantly tilted for $z_c\neq0$. The deviation $\Delta\theta$ of the eigenaxes depends more on $z_c$ than on $x_c$, and reads:
\begin{equation}
\Delta\theta = -\frac{4\xi^2-1}{3(\xi^2-1+3\varepsilon)}\left(\frac{\sqrt{\varepsilon}}{\sqrt{1-4\varepsilon}}\frac{x_c}{x_0}+\frac{z_c}{x_0}\right).
\label{eq:thetashift}
\end{equation}

The dependence of the oscillation frequencies on the plug position is also calculated analytically at first order. A remarkable feature is that the highest frequency $\omega_{x'}$ does not depend on the plug position at first order in $x_c$, $z_c$. This makes this frequency particularly stable. The frequency $\omega_y\propto r_0^{-1/2}$ depends only on $x_c$, as does $r_0$. We express the relative shifts of $\omega_{z'}$ and $\omega_y$ as functions of the unperturbed position $x_0$ of a centered plug given by Eq.~\eqref{eq:x0z0}:
\begin{subequations}
  \begin{eqnarray}
   \frac{\Delta \omega_{x'}}{\omega_{x'}} &=& 0,\\
   \frac{\Delta \omega_y}{\omega_y} &=& -\frac{1}{2}\frac{1}{\sqrt{1-4\varepsilon}}\frac{x_c}{r_0} = -\frac{1}{2}\frac{x_c}{x_0},\label{eq:omegayshift}\\
   \frac{\Delta \omega_{z'}}{\omega_{z'}} &=& -\frac{2}{3}\frac{1-\varepsilon}{1-4\varepsilon}\,\frac{x_c}{x_0} -\frac{4}{3}\frac{\sqrt{\varepsilon}}{\sqrt{1-4\varepsilon}}\, \frac{z_c}{x_0}.\label{eq:omegazshift}
  \end{eqnarray}
  \label{eq:omegashift}
\end{subequations}
From these equations, we conclude that the trap frequencies depend essentially on $x_c$ and only slightly on $z_c$, whereas $z_c$ essentially affects the direction of the trap eigenaxes.

We can evaluate the dipolar and parametric heating rates from this analysis. We find again that dipolar excitation dominates. To limit the heating rate below 1\,nK$\cdot$s$^{-1}$, the pointing noise of the laser at the trap frequencies should be kept below $S_{x} = -70$\,dB$\cdot\mu$m$^2\cdot$Hz$^{-1}$ in the $x$ direction, which gives a constraint three times stronger than in the $z$ direction. In our experiment, a residual heating of 80\,nK$\cdot$s$^{-1}$ is present and is directly linked to measured beam pointing fluctuations of order $S_{x} = -50$\,dB$\cdot\mu$m$^2\cdot$Hz$^{-1}$; see the Appendix. As we will see in the next section, this moderate heating does not prevent the formation of a BEC in the optically plugged trap.

\section{Evaporation to a BEC}
\label{sec:evaporation}

The present section is devoted to the experimental study of the evaporative cooling dynamics of a $^{87}$Rb cold gas in the optically plugged quadrupole trap described in the previous section. The atoms are prepared in the $F=1$, $m_F=-1$ internal ground state.
We first concentrate on Majorana losses in a bare quadrupole trap and show that the simple model introduced in Ref.~\cite{Petrich1995} provides a reasonable description of our experimental observations up to a dimensionless geometrical factor that we are able to measure. Then, careful measurements of the gas phase space density during the evaporation in the quadrupole trap allow us to see that the cooling dynamics can be maintained in the runaway regime down to a temperature where Majorana losses start to prevail. We finally show how to optimize the effect of the plug beam by modifying the magnetic field gradient, to efficiently suppress Majorana losses and reach the BEC threshold.

\subsection{Measuring Majorana losses}

The Majorana loss rate $\Gamma_m$ of a dilute gas confined in a bare quadrupole trap at temperature $T$ can be defined as~\cite{Petrich1995,Heo2011}
\begin{equation}
\Gamma_m=\chi\frac{\hbar}{M}\left(\frac{\hbar\alpha}{k_B T}\right)^2,
\label{eqn:majorana}
\end{equation}
where $\chi$ is a dimensionless geometrical factor and $k_B$ the Boltzmann constant.
Following the approach described in Ref.~\cite{Chicireanu2007}, one can also show that the thermodynamics of the trapped cloud is determined by the following set of equations:
\begin{subequations}
  \begin{eqnarray}
    \frac{\dot{T}}{T}&=&\frac{4}{9}\Gamma_m,\label{eqn:dotT}\\
    \frac{\dot{n_0}}{n_0}&=&-\Gamma_b-\frac{7}{3}\Gamma_m,
  \end{eqnarray}
  \label{eqn:dynamics}
\end{subequations}
where $n_0$ is the gas peak density and $\Gamma_b$ the one-body loss rate due to collisions with the background gas.
The temperature variation rate in Eq.~\eqref{eqn:dotT} is deduced from the loss rate of Eq.~\eqref{eqn:majorana} and from the average energy of atoms lost in a spin-flip event. This last average is computed under the same assumptions that led to Eq.~\eqref{eqn:majorana} and gives an average energy of $(5/2)k_B T$ per lost atom. Using the energy conservation relation $\dot{E}=(5/2)k_BT\dot{N}$ and the expression of the total energy in a quadrupole trap $E=(9/2)k_BNT$, it is then straightforward to obtain Eqs.~\eqref{eqn:dynamics}.

Solving Eqs. \eqref{eqn:dynamics} explicitly, one can show that $n_0$ decays according to a non-exponential law involving both $\chi$ and $\Gamma_b$. In comparison, $T$ has a simpler behavior, where only $\chi$ enters. Denoting by $T_0$ the gas initial temperature, the variation of $T$ is written
\begin{subequations}
  \begin{eqnarray}
  &&T(t) =\sqrt{T_0^2+\gamma t}\, , \label{eqn:MajoranaHeating} \label{eq:temp_inc}\\
  &&\mbox{where} \quad
  \gamma =\frac{8}{9}\chi\frac{\hbar}{M}\left(\frac{\hbar\alpha}{k_B}\right)^2.
  \label{eqn:gamma}
  \end{eqnarray}
\end{subequations}
In order to obtain the value of $\chi$ experimentally, it hence appears more favorable to measure the gas temperature increase rather than the atom number decay, as it involves fewer free parameters. The heating of the gas can be intuitively understood as an anti-evaporation process: the coldest atoms of the gas are indeed the closest to the quadrupole magnetic field zero and thus more likely to undergo a spin flip.

In order to check the validity of this model and determine the value of $\chi$
for our setup, we performed the following experiment: After evaporatively
cooling the atomic cloud down to $T=20\,\mu$K in a bare quadrupole trap with
$b'=216$~G$\cdot$cm$^{-1}$, corresponding to the maximal value of the magnetic gradient in our setup, we decompress the trap in 50~ms to a final value of the magnetic gradient $b'_f$ and keep the cloud in the trap for various holding times. The gas density is then measured by absorption imaging after a time of flight of 12~ms, allowing us in turn to deduce $T$ (see Sec.~\ref{sec:tof}).  

\begin{figure}[t]
\begin{center}
\includegraphics{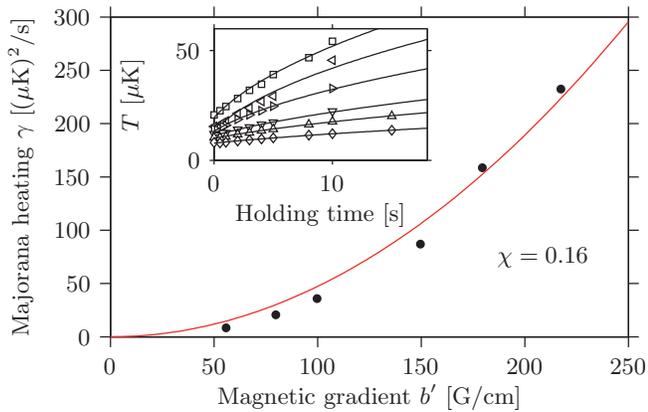}
\caption{(Color online) Measurement of the Majorana heating rate. Parameter
$\gamma$ [see Eq. \eqref{eqn:gamma}] as a function of the trap magnetic field 
gradient. The solid line is a quadratic fit allowing extraction of the geometrical
factor $\chi=0.16$, according to Eq. \eqref{eqn:gamma}. Inset: Cloud temperature
as a function of time, for different magnetic field gradients, fitted by formula
\eqref{eqn:MajoranaHeating} (solid lines). See text for details.}
\label{fig:Majorana}
\end{center}
\end{figure}

The results obtained are presented in Fig.~\ref{fig:Majorana}, and we can see in the inset that the gas temperature increase for different values of $b'_f$ is well fitted by Eq.~\eqref{eq:temp_inc}. Moreover, we correctly recover the quadratic dependence of the heating coefficient $\gamma$ on the magnetic gradient $b'_f$ expected from Eq.~\eqref{eqn:gamma} and deduce $\chi=0.16$. This value agrees well with the one measured in Ref.~\cite{Heo2011} for a gas of $^{23}$Na atoms. This tends to prove that the simple model presented here describes correctly the physics of Majorana losses in a quadrupole magnetic trap, independently of the atomic species.

\subsection{Runaway evaporation in a bare quadrupole trap}

Evaporative cooling consists in truncating the highest part of the energy
distribution of a trapped gas to force it to thermalize at a smaller temperature
than initially. To describe the thermodynamics of the gas during the process, one
usually introduces the truncation parameter $\eta$, which can be defined as the
ratio of the trap depth, fixing the maximal energy of the trapped atoms, to the
gas temperature. It has been shown in~\cite{Luiten1996} that in the absence of
Majorana losses, the efficiency of evaporative cooling is completely determined
by the ratio $r=\Gamma_b/\Gamma_c$ of the background loss rate $\Gamma_b$ to the
elastic two-body collision rate $\Gamma_c$. The authors of~\cite{Luiten1996}
have also shown that below a certain value of $r$ depending on the trap geometry
and on $\eta$, assumed here to be constant, the evaporative cooling dynamics enters
the runaway regime, where $\Gamma_c$ keeps increasing during the evaporation.
For sufficiently small $r$, $\Gamma_c$ and the phase-space density are even expected to diverge in finite time, in turn allowing the gas to efficiently reach the BEC threshold.

\begin{figure}[!t]
\begin{center}
\includegraphics{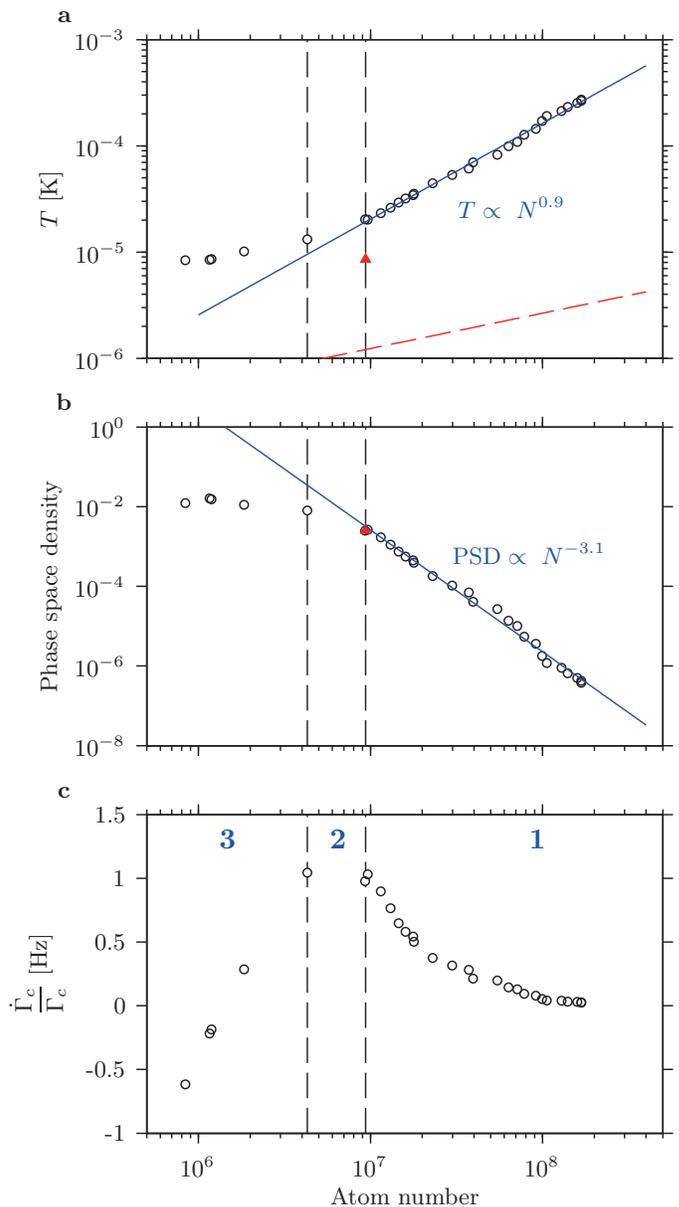}
\caption{(Color online) Evaporation in the bare quadrupole trap.
Black points: experimental data for different trap depths (final frequency) at
the maximal radial gradient of 216\,G$\cdot$cm$^{-1}$. Red triangle: data taken immediately after the trap opening (see text) for a gradient of 55.4\,G$\cdot$cm$^{-1}$.
(a) Temperature versus atom number in log-log scale.
The blue solid line is a linear fit to the data, giving the scaling $T\sim
N^{0.9}$. The dashed red line is the critical temperature expected for a
harmonic trap with the oscillation frequencies given in Table~\ref{tab:param}.
(b) Phase-space density versus atom number in log-log scale, computed from the
measured temperature and the exact knowledge of the potential shape.
(c) Expected $\dot{\Gamma}_c/\Gamma_c$ ratio versus the atom number in
linear-log scale, as computed from the right-hand side of Eq.~\eqref{eqn:Gc}.
On the three graphs, the dashed vertical lines delimit three regions labeled
\textbf{1}, \textbf{2}, and \textbf{3} (see text for details).
}
\label{fig:evap}
\end{center}
\end{figure}

Majorana losses can be expected to deeply modify the evaporative cooling dynamics and in any case reduce its efficiency, especially since $\Gamma_m$ increases with smaller temperatures. Yet maximizing $\Gamma_c$ throughout the evaporation process still appears as the best strategy to be able to stay in the runaway regime during the process. This is achieved in our experiment by optimizing the evaporative cooling ramp shape in the bare magnetic quadrupole trap:  the potential depth is modified through a radio-frequency knife whose frequency $\nu_{\rm rf}$ decreases during the evaporation from $50$ to $4\,$MHz. In order to maximize $\Gamma_c$, the ramping speed $\dot{\nu}_{\rm rf}$ is increased in four steps from $3.6$ to $6.7\,$MHz/s. Different thermodynamical quantities of the gas measured along the resulting evaporative cooling ramp are presented in Fig.~\ref{fig:evap}. 

Interestingly, we see in Fig.~\ref{fig:evap}(a) that the gas temperature during the first part of the evaporation (region \textbf{1}) scales as $T\sim N^{0.9}$ : this value is close to the maximal theoretically predicted exponent of $1.2$ in a linear trap~\cite{Ketterle1996a}. At the same time, the phase-space density scales as ${\rm PSD}\sim N^{-3.1}$ as shown in Fig.~\ref{fig:evap}(b). This result demonstrates the efficiency of the evaporative cooling process and illustrates the weak influence of Majorana losses at these relatively high temperatures. Note that here the truncation parameter $\eta$ keeps an almost constant value of 9.

As soon as $T\leq 20~\mu$K, we observe a sudden decrease of the efficiency of the evaporative cooling, and the trapped gas never reaches the BEC threshold. To illustrate this effect we have plotted in Fig.~\ref{fig:evap}(c) the ratio $\dot{\Gamma_c}/\Gamma_c$, which gives the evolution of the two-body collision rate during the evaporation. To compute this quantity we rely on the simple model introduced in Ref.~\cite{Luiten1996}, which, assuming $\eta\gg1$, predicts
\begin{equation}
  \frac{\dot{\Gamma}_c}{\Gamma_c}=-\Gamma_b-\frac{19}{9}\Gamma_m+f(\eta)\Gamma_c \, ,
  \label{eqn:Gc}
\end{equation}
where $f(\eta)=\frac{2}{9}\left[563+4\eta(5\eta-54)\right]\, e^{-\eta}$. Experimentally the value of $\dot{\Gamma_c}/\Gamma_c$ is estimated from the right-hand side of Eq.~\eqref{eqn:Gc}. We see in Fig.~\ref{fig:evap}(c) that while $\dot{\Gamma_c}/\Gamma_c$ increases faster and faster in region~\textbf{1}, as expected in the runaway regime, it saturates in region~\textbf{2} before quickly decreasing and reaching negative values in region~\textbf{3}. This study illustrates how Majorana losses break down the evaporation efficiency and prevent the gas from reaching quantum degeneracy.

\subsection{Transfer to the optically plugged trap}

In order to circumvent the limitations imposed by Majorana losses, it is necessary to add the optical barrier induced by the plug beam at the  center of the quadrupole magnetic trap. This effectively decreases the atomic density at the  center of the quadrupole field,
resulting in an exponential suppression of Majorana losses~\cite{Heo2011}.

In our experiment, however, simply adding the plug beam to the bare quadrupole magnetic trap described in the previous section is insufficient to reduce Majorana losses enough to allow the gas to reach quantum degeneracy.
Indeed, the light shift is of order $100\,\mu$K, resulting in a potential barrier of only $50\,\mu$K for a gradient of $b'=216\,$G$\cdot$cm$^{-1}$, to be
compared to the 20\,$\mu$K cloud temperature at the point where the evaporation dynamics slows down.

We find that it is necessary to adiabatically open the trap just before the gas
reaches $T=20~\mu$K. This is done by reducing the magnetic gradient $b'$ to $55.4\,$G$\cdot$cm$^{-1}$ in $50\,$ms. The opening has two effects: reducing the temperature to $8\,\mu$K and increasing the barrier to $90\,\mu$K, which effectively suppresses Majorana losses. The resulting lifetime of the atoms in the opened and optically plugged trap reaches $\geq 20$~s. After such a sequence, $r$ is reduced to $\sim10^{-3}$ which is small enough to carry on the evaporative cooling.

As discussed in Sec.~\ref{sec:noise}, the magnetic field value $B_0$ at the trap minimum, or equivalently the trap bottom frequency $\nu_0=\mu B_0/h$, may be adjusted by a controlled shift of the plug along the $x$ axis, as $\Delta \nu_0/\nu_0 = x_c/x_0$; see Eq.~\eqref{eq:r0ofxc}. The subsequent relative change $\Delta\bar{\omega}/\bar{\omega}$ in the mean trapping frequency $\bar{\omega}=(\omega_{x'}\omega_y\omega_{z'})^{1/3}$, and thus in the critical temperature, is about three times smaller. This gives room for a possible adjustment of the bottom frequency alone. Indeed we are able to produce quasi-pure BECs with trap bottom frequencies ranging from $\sim100\,$kHz to $\sim300\,$kHz, by logarithmically sweeping $\nu_{\rm RF}$ from $2\,$MHz down to $\nu_0+50\,$kHz in $5\,$s (see the Appendix for details). Let us note that the smooth dependence of $\bar{\omega}$ with respect to $x_c$ and $z_c$ ensures that small long-term drifts in the plug position do not change much the experimental conditions needed to reach the condensation threshold, apart from a shift in the final evaporation frequency.

The results presented in this section show that the Majorana losses can be precisely measured during the evaporation process and account for the observed breakdown in evaporation dynamics. The temperature at which this occurs sets the order of magnitude of the plug barrier necessary to suppress Majorana losses, and an adiabatic trap opening might be required depending on the available laser power. Let us note that in our setup we are never limited by three-body collisions, due to relatively low atomic densities. However, it has been shown~\cite{Heo2011} that trap opening also provides a way to circumvent the three-body losses, and thus produce large BECs in optically plugged traps.

\section{Characterization}
\label{sec:characterization}

The hybrid optical and magnetic final trap is close to a linear trap at energies larger than $20\,\mu$K and is harmonic at very low energies, below $1\,\mu$K. Between these two regimes, it is strongly anharmonic, and its precise features depend on several parameters: the magnetic field gradient $b'$, the waist and power of the plug beam, and its position with respect to the magnetic field zero. As described in Sec.~\ref{sec:BEC}, the trap characteristics are mostly sensitive to the plug beam position. The beam position determines the magnetic field $B_0$ at the bottom of the trap, the oscillation frequencies, and the population ratio between the two potential minima on both sides of the laser. In this section, we give a series of measurements that allow a full characterization of the trap parameters, and compare them to calculations of the potential.

\subsection{rf spectroscopy}
\label{sec:spectro}
First, we recall that the oscillation frequency $\omega_y$, given by the expression \eqref{eq:omegay}, depends only on the magnetic gradient $b'$ and the magnetic field at the trap bottom $B_0$. The gradient $b'$ is known by both a calibration of the quadrupole coils with a Hall probe and a direct measurement with a cold cloud; see Appendix~\ref{sec:overview}. A measurement of $B_0$ would thus allow a prediction of the frequency $\omega_y$.

The frequency $\nu_0$ corresponds to the rf frequency at which all the atoms are evaporated from the trap. $B_0$ can thus be estimated directly from the evaporation procedure. However, the rf amplitude used for evaporative cooling is large enough to ensure an adiabatic deformation of the magnetic level, and thus shifts the bottom frequency. To get a precise value of the rf frequency at the bottom of the trap, we perform radio-frequency spectroscopy of trapped condensates.
We use an additional radio-frequency field linearly polarized along the $y$ axis to resonantly probe the trapped atoms.
This induces spin flips between the $F=1$, $m_F=-1$ trapped state and the $F=1$, $m_F=0$ state in which the atoms are pushed away by the combination of the plug potential and gravity. We extract from the spectroscopy data the resonant rf field at the trap bottom.

We calibrate the rf probe by monitoring the atom number decay for small rf amplitudes at resonance. The decay rate is found to scale quadratically with the rf Rabi frequency, as expected from a Fermi golden rule estimation. We follow the approach of Ref.~\cite{Gerbier2001} to compute the decay rate as a function of the trap parameters.
Although the purely magnetic trap used in Ref.~\cite{Gerbier2001} is completely different from ours, the trap shape presents similarities. Indeed, in Ref~\cite{Gerbier2001}, the vertical trapping frequency is so small that the entire atomic cloud is displaced by the gravitational field to a region where the atoms experience a magnetic field gradient. Similarly, in our setup the atoms are maintained in a region with a magnetic field gradient by the effect of the plug. In that sense, the magnetic landscape is the same in these two different traps and Eq. (7) of Ref. \cite{Gerbier2001} is still valid in our setup, even if the predicted spectrum width is more complicated to estimate, given the tilt of the eigenaxes. Thus we use both the measured decay rates and the measured spectral widths to calibrate the Rabi frequency.

\begin{figure}[t]
\begin{center}
\includegraphics{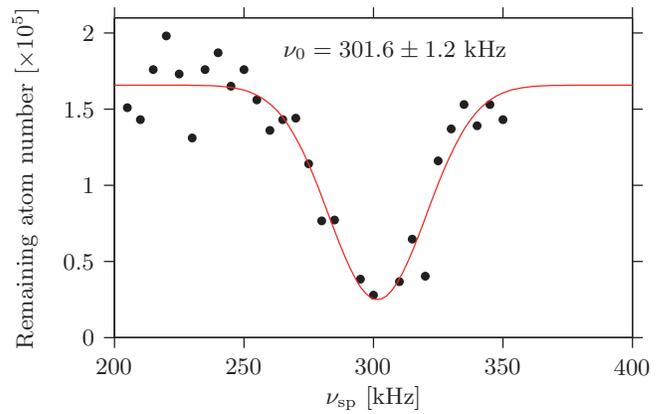}
\caption{(Color online) rf spectroscopy: Number of atoms remaining in the trap after a 100\,ms rf pulse, as a function of the rf frequency. The solid line is a Gaussian fit to the data.}
\label{fig:spectro_RF}
\end{center}
\end{figure}

Typical spectra are obtained with about $100\,$Hz of rf Rabi frequency, as shown in Fig.~\ref{fig:spectro_RF}. To avoid projection losses when the rf is turned on and off, we increase (decrease) linearly its amplitude in $1\,$ms and probe the cloud during $100\,$ms. As expected Fig.~\ref{fig:spectro_RF} shows a symmetric spectrum with a  center frequency of $\nu_0=301.6\pm1.2\,$kHz. As seen in Sec.~\ref{sec:BEC}, the oscillation frequency in the $y$ direction depends only on the cloud's effective distance from the quadrupole  center $r_0=2\pi\nu_0/\alpha$ [see Eq.~\eqref{eq:omegay}]. This measurement of the trap bottom frequency thus gives  $r_0 =77.8\pm 1\,\mu$m and the resulting value of the oscillation frequency $\nu_y=\omega_y/(2\pi)=76.2\pm 0.8\,$Hz. The uncertainty on $\nu_y$ mainly comes from the uncertainty on the magnetic gradient; see Table~\ref{tab:param}.

\subsection{Measurement of the oscillation frequencies}
\label{sec:measOsc}

The most straightforward method of reliably measuring the oscillation frequencies of a harmonic trap consists in exciting dipolar and/or parametric oscillations of the trapped cloud. In the case of an optically plugged trap this can be achieved by modulating the current $I$ in the coils which produce the magnetic quadrupole. In this way, both the cloud position and the trap frequencies are modulated~\cite{note2}.

In our experiment we impose $I(t)=I_0(1+\epsilon\sin{2\pi\nu_{\rm ex} t})$, where $I_0$ is the mean current amplitude, $\epsilon=1.8\,\%$ is the amplitude of the modulation and $\nu_{\rm ex}$ is the excitation frequency. After a trap modulation of 500\,ms, the excitation is converted into heating during an additional 100\,ms thermalization time. We then switch off the confinement abruptly. The atoms are finally observed after a 25\,ms time of flight (TOF) and the rms size of the cloud $s_z(t_{\rm tof})$ is recorded.

\begin{figure}[t]
\begin{center}
\includegraphics{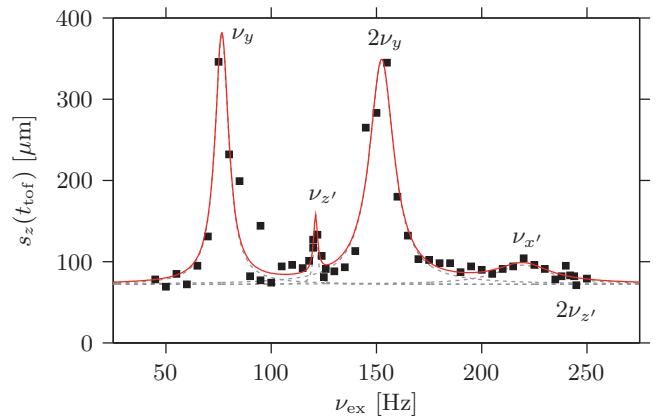}
\caption{(Color online) Spectroscopy of the oscillation frequencies. Squares: Measured cloud size in micrometers after 25\,ms time of flight, as a function of the modulation frequency of the current in the quadrupole coils. Solid red line: Multiple Lorentzian fit to the data.}
\label{fig:spectro_freq}
\end{center}
\end{figure}

In Fig.~\ref{fig:spectro_freq} we display the dependence of $s_z(t_{\rm tof})$ on $\nu_{\rm ex}$ measured experimentally. The different peaks in the graph correspond to the dipolar and parametric resonances of the trap. Their position is deduced from a Lorentzian fit. The first peak, centered at 76.6(4)~Hz, directly corresponds to $\nu_y$ which nicely confirms the estimation of the previous section. We otherwise find $\nu_{x'} = 220(10)$\,Hz\ and $\nu_{z'} = 121.1(5)$. The uncertainty results from the fit. Note that parametric resonances happen at double the frequency of dipolar resonances.

\subsection{Trap parameters}
From the knowledge of the oscillation frequencies, we can infer the trap parameters. The magnetic gradient $b'$ is measured independently to be $55.4\pm0.6$\,G$\cdot$cm$^{-1}$ (see the Appendix~\ref{sec:overview}) which means that both $\alpha$ and $\varepsilon = 0.0337\pm0.0007$ are known.  The laser power can also be measured in a reliable way to be $P=5.8\pm0.1\,$W by recording the power before and after the vacuum chamber. The three remaining independent parameters, $w_0$, $x_c$, and $z_c$, which are more difficult to obtain by a direct measurement, are deduced from the three measured oscillation frequencies.

The value of $\omega_y$ directly gives $r_0$, which is the isomagnetic surface on which the minimum lies, from Eq.~\eqref{eq:omegay}, which also holds for an off-centered plug beam. The same information also comes from the rf spectroscopy. Now, the general idea for finding the trap parameters is to first adjust the tentative value of $w_0$ to fit the correct $\omega_{x'}$. As $\omega_{x'}$ does not depend on the plug position at first order, the expression for $\xi$~\cite{note1}, which enters in $\omega_{x'}$, can be safely used. Let us introduce the light shift $u_0 = U_0/(\hbar\alpha w_0)$ in dimensionless units. $u_0$ is a function of $P$ and $w_0$: $u_0=k P/w_0^3$, where $k$ is a fixed parameter for a given gradient and is equal to $k=1.98\times 10^5$\,W$^{-1}\cdot\mu$m$^3$ in our case with the gradient of Table~\ref{tab:param}. The function $\xi(w_0)$ is thus known for the laser power $P=5.8$\,W, which allows us to determine $w_0$ from $\omega_{x'}$ by inverting the relation
\begin{equation}
\omega_{x'} = (1-3\varepsilon)^{1/4}\,\sqrt{\frac{4\xi^2(w_0) - 1}{\xi(w_0)}}\,\sqrt{\frac{\hbar\alpha}{Mw_0}}.
\label{eq:omegaxofw0}
\end{equation}
We find $w_0=46\,\mu$m, in fair agreement with an estimate from a direct optical measurement.

The knowledge of $w_0$ allows prediction of a zero-order value $r_0^0$ for the effective radius from Eq.~\eqref{eq:r0}. From the shift $\Delta r_0 = r_0 - r_0^0$ of this value compared with the measured one, we deduce the horizontal plug shift $x_c = \sqrt{1-4\varepsilon}\Delta r_0 $; see Eq.~\eqref{eq:r0ofxc}. Finally the value of $z_c$ is chosen to fit $\omega_{z'}$, a first guess being given by Eq.~\eqref{eq:omegazshift} where the zero-order frequency comes from Eq.~\eqref{eq:omegaz}. Application of this method allows us to determine the values of $w_0$, $x_c$, and $z_c$ given in Table~\ref{tab:param} and calculate the trap depth. The uncertainties are deduced from the relations \eqref{eq:omegaxofw0}, \eqref{eq:r0}, and \eqref{eq:omegashift} and from the uncertainties of the experimental measurements.

\subsection{Time-of-flight analysis}
\label{sec:tof}

\begin{figure}[t]%
	\centering
	\includegraphics{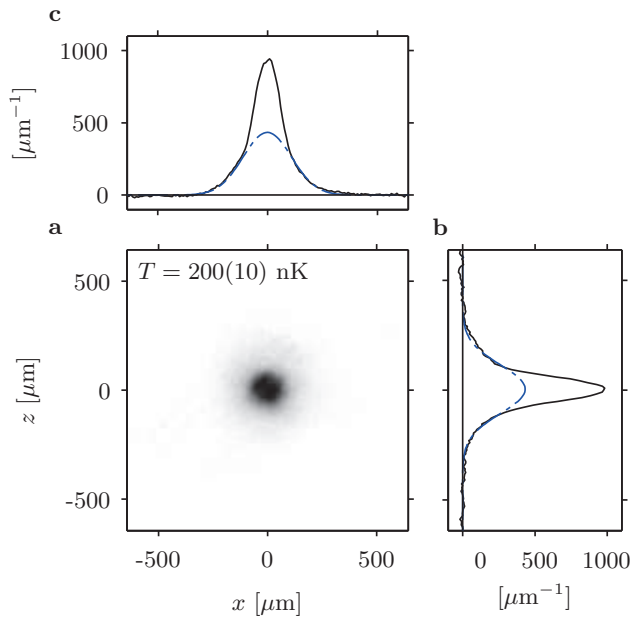}
	\caption{(Color online) Analysis of time-of-flight image. \textbf{(a)} Absorption image of a cold atomic cloud after 25~ms time of flight. High optical density at the center of the image indicates the presence of a BEC. Thermal tails are yet still visible. \textbf{(b)} Integrated profile of \textbf{(a)} along the $x$ axis. The dashed blue line corresponds to the integrated profile of a two-dimensional (2D) fit of the thermal tail of the atomic cloud displayed in \textbf{(a)}. \textbf{(c)} Same as \textbf{(b)} with integration along the $z$ axis. }%
	\label{fig:tof}
\end{figure}

In order to access the temperature of the trapped atomic clouds we rapidly switch off the confinement ($<0.5$\,ms) and let the gas expand during a controllable duration $t_{\rm tof}$. Analysis of the shape of the atomic column density obtained from absorption imaging (see the typical data displayed in Fig.~\ref{fig:tof}) allows us to deduce the temperature of the trapped gas. Depending on the physical regime of the cloud, different methods are used.

A simple strategy to obtain the temperature of the gas consists in measuring the rms width of the cloud $\sigma_u$, with $u=x,y,z$.  For a freely expanding gas it can be expressed as
\begin{equation}\label{eq:rms_time}
\sigma_u^2 \left( t_{\rm tof} \right)=\sigma_u^2\left(0 \right)+\left[\sqrt{\frac{k_B T}{m}}t_{\rm tof} \right]^2,
\end{equation}
where
\begin{equation}\label{eq:rms_in_situ}
\sigma_u^2\left(0 \right)=C \int d\mathbf{r} \ u^2 \exp[-\beta U\left(\mathbf{r}\right)]
\end{equation}
with $\beta=\left[k_B T\right]^{-1}$ and $C=\left[ \int d\mathbf{r} \ \exp[-\beta U\left(\mathbf{r}\right)] \right]^{-1}$. For sufficiently high temperatures, the effect of the optical plug on the trapped gas can be neglected and $U\left(\mathbf{r}\right)$ replaced by $U_B\left(\mathbf{r}\right)+Mgz$ in Eq.~\eqref{eq:rms_in_situ}. The calculation of $\sigma_u^2\left(0 \right)$ then becomes analytical and using Eq.~\eqref{eq:rms_time} it is straightforward to deduce $T$ from a single picture. As soon as the gas gets too cold, approximating $U\left(\mathbf{r}\right)$ by $U_B\left(\mathbf{r}\right)+Mgz$ is no longer valid and $\sigma_u^2\left(0 \right)$ can be obtained only numerically. At this point, a better strategy to deduce $T$ consists in taking a series of images with different $t_{\rm tof}$ and use Eq.~\eqref{eq:rms_time}.

Below the Bose-Einstein condensation threshold, the expansion of the gas cannot be considered as ballistic and we have to rely on a different strategy. We hence fit an analytical formula based on ideal Bose gas theory~\cite{Gomes2006} to the thermal tail of the atomic density distribution and deduce the value of $T$. In principle, such an approach could be extended to the investigation of the BEC parameters, like the initial Thomas-Fermi radii~\cite{Castin1996}. However, technical limitations of the imaging system have prevented us from applying this analysis.

\section{Conclusion}
\label{sec:conclusion}

In this paper, we describe the successful evaporation of a rubidium 87 gas to Bose-Einstein condensation in a linear quadrupole trap plugged by a laser beam at 532\,nm. We provide a simple model which describes the trap characteristics. We check its predictions quantitatively by a direct measurement of the oscillation frequencies. For example, the value of $\omega_{z'}/\omega_y=1.61$ predicted by the simplified model with a centered plug [see Eq.~\eqref{eq:omegaz}] agrees with a direct measurement within 2\%.

We present two spectroscopic tools to fully characterize the trap. Other possible diagnostics include the separation between the two minima, measured by \textit{in situ} absorption imaging with the laser and the magnetic field on (see the inset of Fig.~\ref{fig:potential}) or the orientation $\theta$ of the eigenaxes measured after a free flight. These methods could improve the error bars on the measured parameters. However, the two proposed methods measure a frequency directly and are thus more reliable as they do not require a particularly high imaging quality. In particular, the measurement of the bottom frequency $\nu_0$ gives access to the main scaling frequency $\omega_y$. Provided an independent estimation of the plug beam waist is available, this single measurement allows a good estimation of the average frequency $\bar{\omega}$, which is the natural scale for the critical temperature or the chemical potential~\cite{PitaevskiiStringari}.

A detailed study of the influence of the plug position on the trap parameters confirms that the trap characteristics are robust with respect to small deviations in the experimental parameters. The only relevant source of heating comes from beam pointing fluctuations. Again, their effects remain modest. Without active stabilization of the pointing, the dipolar heating rate is kept to a reasonable value of 80\,nK$\cdot$s$^{-1}$ in our experiment; see the Appendix. 

Evaporation in the optically plugged quadrupole trap is shown to lead efficiently to degeneracy, with a dynamics comparable to evaporative cooling of rubidium in similar linear traps~\cite{Lin2009}. Another result of the paper is the measurement of the effective volume entering in the model of Majorana losses~\cite{Petrich1995}. We find a value in agreement with recent measurements in a similar setup with sodium atoms~\cite{Heo2011}.

Finally, we have shown that a quadrupole trap optically plugged with a 532\,nm focused beam, initially demonstrated with sodium atoms~\cite{Davis1995}, is also well adapted to the production of degenerate rubidium gases, despite the much larger detuning. It should also work as well with atoms with intermediate wavelengths for the main dipole transition, like lithium (671\,nm) or potassium (767\,nm). This makes this trap particularly well suited to production of mixtures of degenerate gases.

\acknowledgments
We acknowledge Institut Francilien de Recherche sur les Atomes Froids (IFRAF) for support. LPL is UMR of CNRS and Paris 13 University. R. D. acknowledges support from an IFRAF grant. We thank Bruno Laburthe-Tolra for helpful comments on the evaporation dynamics.

\appendix
\section{Experimental setup}
\label{sec:experiment}

In this appendix we first give an overview of the different elements of the experimental setup. Then we describe the laser sources and finally give technical details on the experimental sequence.

\subsection{Overview of the experimental setup}
\label{sec:overview}
\begin{figure}[b]%
	\centering
 	\includegraphics[width = 0.8\linewidth]{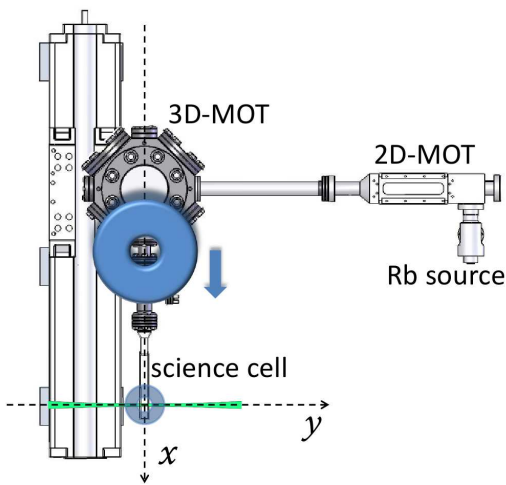}%
	\caption{(Color online) Scheme of the experimental setup. The three main chambers (2D MOT, octagonal chamber for the 3D MOT and science cell) are visible, as well as the rail for the transport of the magnetic coils (on the left). The MOT beams propagate at $45^{\circ}$ with respect to the $x$ and $y$ axes. The moving coils are represented by a blue hollow disk. The plug beam (in green) is aligned with the $y$ axis. The $x$ and $y$ axes cross at the  center of the quadrupole trap. The remaining part of the vacuum chamber (ionic pumps, etc.) have been omitted for clarity.}%
	\label{fig:setup}
\end{figure}

The experimental setup can be decomposed into three main parts as depicted in Fig.~\ref{fig:setup}. A first vacuum chamber is filled with a hot vapor of $^{87}$Rb atoms heated in an oven to $70\,^\circ$C. This sets the $^{87}$Rb partial pressure to $10^{-8}$\,mbar. The atoms are collected by a bi-dimensional magneto-optical trap (2D MOT)~\cite{CheinetThese}. An additional laser beam pushes the atoms into a second vacuum chamber where they are captured by a three-dimensional magneto-optical trap (3D MOT). We typically load about $2.5\times 10^9$ atoms in the 3D MOT in 5\,s, these numbers being mostly sensitive to the power and direction of the pushing laser beam. Thanks to a differential pumping tube the partial pressure in the 3D MOT chamber can be kept of the order of $10^{-10}$\,mbar, as deduced from the 3D MOT lifetime of 30\,s~\cite{Steane1992}.

One of the main features of our setup comes from the water-cooled 3D MOT quadrupole magnetic field coils which are held on a motorized translation stage that can be displaced at will along the $x$~axis~\cite{Lewandowski2003}. After the transfer of the atoms from the 3D MOT into a quadrupole magnetic trap made with the same coils, the translation stage is moved along a 28.5\,cm path, bringing in turn a fraction of the atoms into a $10\times 10$\,mm$^2$ inner-size science glass cell (Starna) with very good optical access. The cell walls have a width of 1.25\,mm and are anti-reflection coated on their external sides at 532\,nm and between 650 and 1100\,nm. During the transport in the moving trap, the atoms pass through a 4-mm-diameter, 94-mm-long tube, which ensures an ultrahigh-vacuum environment in the final chamber. The atoms are finally transferred into a second quadrupole magnetic trap induced by a pair of water-cooled conical coils made of $40$ loops of hollow copper tubes
  each. There the loss rate $\Gamma_b$ due to collisions with the background gas corresponds to a lifetime $\Gamma_b^{-1}=120$\,s.

The calibration of the magnetic gradient is done directly with an ultracold cloud, in the following way. The magnetic field is suddenly switched off for a short duration, such that the atomic cloud starts to fall in the gravitational field. It is then switched on again, and the atoms oscillate along the $z$ vertical fiber axis in both the quadrupole and gravitational fields. The two gradients add when the cloud is above the magnetic zero and subtract below. From a parabolic fit to the data and the knowledge of $g$, we infer a calibration of the imaging system and the value of the magnetic gradient. 

\subsection{Laser system}
All the 780 nm light sent onto the atoms is prepared on a separate optical table and carried through single-mode polarization-maintaining fibers.

For cooling and pushing Rb atoms in the 2D and 3D MOTs, we built an agile and powerful 780 nm laser source relying on frequency doubling of an amplified Telecom laser~\cite{Thompson2003}. A fiber-pigtailed, 2\,mW, distributed-feedback laser diode (Fitel) emitting at 1560\,nm feeds a 40-dB erbium doped fiber amplifier (Keopsys), with a maximum output power of 10\,W. The second harmonic is generated in a $3\times 0.5\times 50$\,mm$^3$ periodically poled lithium niobate (PPLN) crystal (HC Photonics). The quasi-phase matching condition is met by regulating the temperature of the crystal at $85\pm 0.1\,^{\circ}$C with a home-made oven. The 1560\,nm beam has a 70\,$\mu$m waist to maximize the doubling efficiency according to the Boyd-Kleinman model~\cite{Boyd1968}. In single pass, we obtain a maximum of 2\,W of linearly polarized light at 780\,nm in a TEM00 mode.

The frequency control of the doubled laser is made by beat note locking with a reference laser. This reference laser is a 780\,nm, 70\,mW,  narrow linewidth external cavity laser (RadiantDye) locked by the saturated absorption technique. Its free linewidth has been measured to $3\,$kHz and its linewidth when locked has been estimated at around $180\,$kHz. The beat note between the reference laser and the doubled laser is frequency locked to an rf signal of adjustable frequency around 270\,MHz.

By tuning the beat-note frequency we are able to sweep the doubled laser frequency over a large span from $+10\Gamma$ to $-60 \Gamma$ around the $5S_{1/2},F=2\rightarrow 5P_{3/2},F'=3$ cycling transition of the $^{87}$Rb $D_2$ line, where $\Gamma$ is the transition linewidth, without altering the output intensity. The intensity is independently controlled or switched off by an acousto-optic modulator. The doubled light is then split into four beams injected into polarization maintaining fibers. Two transverse cooling beams and a weak pushing beam, with a total power of 120\,mW, seed the 2D MOT. The last one is used for the 3D MOT and is coupled to one of the two input ports of a $2\times 6$ fiber cluster (Sch\"after+Kirchhoff).
  
At the cluster output, each of the six fibers is connected to a compact three-lense collimator system (SYRTE Labs design). The collimated beams are clipped to a diameter of 1 in. by a quarter-wave plate directly set at the collimator output to produce the required circularly polarized light. The total intensity of the six cooling beams is 41\,mW$\cdot$cm$^{-2}$.

The repumper laser is a 70\,mW Sanyo laser diode, frequency locked to the $F=1\rightarrow F'=2$ transition of the $D_2$ line. It is superimposed onto the 2D MOT transverse beams before their injection into the fibers, while it is mixed with the 3D MOT beams through the second input of the fiber cluster. The repumper laser is not present in the pushing beam, which limits the 2D MOT atomic beam velocity and improves the recapture efficiency~\cite{Dimova2007}.

The plug beam originates from a high-power diode-pumped laser at 532\,nm (Spectra Physics Millennia X) with an output power reaching 10\,W in a TEM$_{00}$ spatial mode. To control the laser intensity and allow for the fast switching of the beam, we use an acoustic-optic modulator with a  center frequency of 110\,MHz. At the position of the atoms, the optical plug power is about 6\,W for a waist of $46\,\mu$m. The beam position is controlled by two piezoelectric actuators on a mirror mount.

The power spectrum density of intensity fluctuations of the laser beam amounts to $-93$\,dB$\cdot$Hz$^{-1}$ at the trap frequencies, which corresponds to a calculated heating rate of about 1\,nK$\cdot$s$^{-1}$; see Sec.~\ref{sec:noise}. Using a quadrant photodiode, we characterized the
pointing stability of the plug beam. The long-term beam pointing stability is
rather good, with a drift below $1\,\mu$m over 1 week. We also recorded the
power spectrum of beam pointing fluctuations. We find
$S_x\leq-50\,$dB$\cdot\mu$m$^{2}\cdot$Hz$^{-1}$ over the trap frequency range. This value agrees with a measured heating rate of 80\,nK$\cdot$s$^{-1}$ in the optically plugged quadrupole trap. This figure could be improved if necessary by actively locking the plug position to a reference measured with the quadrant photodiode. Finally, the waist fluctuations typically take place at very low frequencies, linked to thermal effects in the laser cavity. We did not quantify this level, but we expect an extremely low value at the trap frequencies, with negligible contribution to the heating rate.

\subsection{Experimental sequence}
\label{sec:exp_sequence}
Before the transfer of the atoms from the 3D MOT to the quadrupole magnetic trap, the 3D MOT is first compressed by progressively increasing the magnetic field gradient from 9.5 to 45\,G$\cdot$cm$^{-1}$ and increasing the 3D MOT laser beams detuning from $-3\Gamma$ to $-14\Gamma$. The laser beams are then turned off and the magnetic field axial gradient is linearly increased from 45 to 200\,G$\cdot$cm$^{-1}$ in 250\,ms, allowing the capture in the magnetic field quadrupole trap of about $10^9$ atoms in the $\left|F=1, m_F=-1\right\rangle$ state. 

The displacement of the quadrupole trap from the 3D MOT chamber to the science glass cell relies on a motorized translation stage (Parker 404-XR) whose maximal velocity, acceleration, and jerk are controlled by built-in software. During the first half of the transportation the maximal acceleration of the translation stage is set to 0.8\,m$\cdot$s$^{-2}$ and its jerk to 50\,m$\cdot$s$^{-3}$. It hence reaches a maximal velocity of 1\,m.s$^{-1}$. The second half of the transfer corresponds to the mirrored image of the first half. Finally the transfer efficiency is about $20\%$ and is mostly limited by the free evaporation of the hottest atoms against the walls of the small tube during the displacement of the translation stage.

As soon as the atoms have reached the science glass cell, the current in the
transfer magnetic field coils is ramped down to zero while the current in the
conical magnetic field coils is ramped up from zero in order to keep the
magnetic field gradients constant. Up to $3\times 10^8$ atoms with a temperature of $150\,\mu$K are loaded in the final magnetic field quadrupole trap. The current is provided by a Delta Elektronika power supply, which has a low relative current noise, $-125$\,dB$\cdot$Hz$^{-1}$ at most, and the associated dipolar heating is negligible. In order to increase the collision rate, the axial magnetic field gradient is finally adiabatically ramped up to 432\,G$\cdot$cm$^{-1}$, the radial gradient thus being $b'=216$\,G$\cdot$cm$^{-1}$. In this trap the atom lifetime is larger than 120\,s at the initial temperature of $270\,\mu$K.

The plug is then switched on and the rf-induced evaporation starts. The plug alignment proceeds as follows. We first leave the plug on during the time of flight and absorption imaging which allows us to image the plug as a depression in the expanding gas density profile. In this way, we are able to roughly match the plug position to the magnetic trap  center (known from \textit{in situ} absorption imaging). Then, at lower temperatures (typically at 20$\,\mu$K), the plug position is optimized to obtain a symmetric expansion of the cloud due to initial acceleration induced by the plug during the time of flight. Once this optimization is done, we proceed with the evaporation sequence and the adiabatic trap opening discussed in Sec.~\ref{sec:evaporation}. Finally we switch off the plug and the magnetic trap simultaneously and optimize the plug position on the peak density after a given time of flight. We repeat this for lower and lower rf frequencies, until the expanded cloud density p
 rofile starts to deviates from a pure Gaussian. At this point BEC is observed on further lowering of the RF frequency. Eventually we finely tune the plug position to place the BEC on a given magnetic isopotential.

Both the rf field applied for evaporation and the probing rf field are linearly
polarized along the $y$ axis. The evaporation field is generated by the rear
panel output of a direct digital synthesizer (DDS) (TaborElec WW1072), going through a $5$\,W amplifier (Kalmus) and roughly impedance matched to a 10-mm-radius 20-loop coil, to avoid any unwanted resonance in this frequency range. The overall circuit length, including rf switches, is much smaller than the shortest wavelength to avoid high-frequency resonances.

The frequency is ramped from 50 down to 4\,MHz with a piece-wise linear time dependence, in order to keep efficient evaporation dynamics in 13.5\,s. At this point we have about $10^7$ atoms at $20\,\mu$K. The trap is then adiabatically opened to a 55.4\,G$\cdot$cm$^{-1}$ gradient in 50\,ms, which results in a cloud temperature of $8\,\mu$K. After the trap opening, the evaporation is carried on with a second antenna (30\,mm radius, 20 loop coil, about 30\,mm away from the trap  center), producing a rf field linearly polarized along the $y$ axis. This antenna is fed directly by a Stanford DS345 function generator, producing a logarithmic ramp from 2\,MHz to 300\,kHz in 5\,s. The probe antenna for radio-frequency probing the condensate is a 30-mm-radius 20-loop coil, directly fed by another Stanford DS345.

We use absorption imaging to measure the density distribution of the atoms once released from the optically plugged quadrupole trap. The probe beam is derived directly from the reference laser, its waist at the atoms position is $1.6\,$mm for a total power of $300\,\mu$W. It is circularly polarized and is superimposed onto the plug beam along the $y$ axis thanks to a dichroic mirror. During the $100\,\mu$s imaging pulse, a bias magnetic field of 1\,G is applied along $y$ to define a quantization axis. After transmission through the science cell, the probe and plug beams are separated with a polarizing beamsplitter (at 532\,nm, transparent and polarization independent for the probe beam at 780\,nm).  The cloud is imaged with a single lens of focal length $250\,$mm onto a CCD camera (Andor IXON-885D) with a pixel size of 8\,$\mu$m. The magnification is 0.94, as measured by the procedure described in Appendix~\ref{sec:overview}. To avoid the saturation of the came
 ra pixels by the remaining optical plug photons, we use a green-light filter (Layertec).


\begin{thebibliography}{24}
\bibitem[{not({\natexlab{b}})}]{note1}
\bibinfo{note}{The parameter $\xi$ can be given explicitly through the $-1$
  branch of the $W$-Lambert function, $W_{-1}(u)$, in the region
  $-\frac{1}{e}<u<0$ and $W_{-1}(u)<-1$: $$\xi =
  \displaystyle\frac{1}{2}\left\{-W_{-1}\left[-\left(\frac{\hbar\alpha
  w_0}{2U_0}\right)^2(1-3\varepsilon)\right]\right\}^{1/2}.$$}.
\bibitem[{not({\natexlab{c}})}]{note2}
\bibinfo{note}{The position of the minimum along $y$ may depend on the magnetic
  gradient $b'$ if a bias magnetic field along $y$ is present, which allows for
  dipolar excitation.}
\end{thebibliography}
\end{document}